\documentclass[prd,aps,showpacs,preprintnumbers,nofootinbib,eqsecnum,preprintnumbers,twocolumn]{revtex4}
\usepackage{eurosym}
\usepackage{amsfonts}
\usepackage{amsmath}
\usepackage{amssymb,epsf}
\usepackage{color}
\usepackage{graphicx}
\usepackage{epstopdf}
\usepackage{float}
\usepackage{caption}
\usepackage{subfig}

\begin{document}

 \title{NED-AdS black holes, extended phase space thermodynamics and Joule--Thomson expansion}

 \author{ S. I. Kruglov\footnote{serguei.krouglov@utoronto.ca}}
  \affiliation{Department of Physics, University of Toronto, 60 St. Georges St., Toronto, ON M5S 1A7, Canada\\
Department of Chemical and Physical Sciences, University of Toronto, 3359 Mississauga Road North, Mississauga, ON L5L 1C6, Canada \\
Canadian Quantum Research Center, 204-3002 32 Ave Vernon, BC V1T 2L7, Canada\\}

\begin{abstract}
We study and discuss the critical behavior of charged by nonlinear electrodynamics (NED) AdS black holes in an extended phase space. The negative cosmological constant is treated as a thermodynamic pressure and the black hole mass is considered as the chemical enthalpy. We show the analogy of thermodynamics and phase transitions studied with the Van der Waals liquid-gas system. We introduce new thermodynamic quantities conjugated to the NED coupling and a magnetic charge. It was demonstrated that the first law of black hole thermodynamics and the Smarr relation hold. We have calculated the Joule–Thomson thermodynamic coefficient and
inversion temperature, and the Joule--Thomson adiabatic expansion are investigated.

\end{abstract}

\maketitle
04.70.-s, 04.70.Bw, 04.20.Dw


\section{Introduction}

Now, it is confirmed that black holes are thermodynamics systems \cite{Bardeen, Jacobson, Padmanabhan} where the black hole area is an entropy and its surface gravity is connected with the temperature \cite{Bekenstein, Hawking}. Next important step is to develop the quantum gravity theory. Anti-de Sitter (AdS) space-time with a negative cosmological constant leads to phase transitions of black holes \cite{Page}. The application of AdS space-time picture is a holography with black holes being a system which is dual to conformal field theories \cite{Maldacena, Witten, Witten1}. This analogy allows us to solve problems of quantum chromodynamics \cite{Kovtun} and condensed matter physics \cite{Kovtun1, Hartnoll}. The cosmological constant can be considered as a pressure that is a conjugate to a volume in an extended phase space black hole thermodynamics. The black hole phase transitions in such picture mimic the liquid-gas phase transitions in ordinary thermodynamics \cite{Dolan, Teo}. There were studies of Born--Infeld electrodynamics coupled to gravity with the negative cosmological constant \cite{Fernando, Dey, Cai, Fernando1, Myung, Banerjee, Miskovic} where an analogy with physics of Van der Waals fluids was shown. In this paper we study and discuss the black hole thermodynamics in the model of nonlinear electrodynamics (NED) \cite{Bronnikov} coupled to gravity in AdS space-time. The global black hole stability and phase transitions are analysed in the extended phase space. The interest to NED model \cite{Bronnikov} considered is due to its simplicity: the metric function is expressed via simple elementary function. This model was explored to study the supermassive black hole M87* \cite{Vagnozzi} and to construct non-singular model of magnetized black hole \cite{Kr}.
There are NED \cite{Soleng, Kr0, Kr1, Krug} (and others) which have attractive features such as the finite electric fields in the center of charges and the finite electrostatic energy that features are similar to Born--Infeld electrodynamics. The extended phase space of AdS black hole
thermodynamics was studied in \cite{Teo, Zou, Hendi, Hendi1, Zeng, Majhi, Jafarzade, Pradhan, Pradhan1, Bhattacharya, Cong, Mann3, Mann1, Mann2}.

Units  $c=\hbar=k_B=1$ are used.

\section{NED-AdS solution}

We start with the action of NED-AdS black holes
\begin{equation}
I=\int d^{4}x\sqrt{-g}\left(\frac{R-2\Lambda}{16\pi G_N}+\mathcal{L}(\mathcal{F}) \right),
\label{2.1}
\end{equation}
where $\Lambda=-3/l^2$ is the negative cosmological constant, $G_N$ is Newton's constant and $l$ is the AdS radius. The NED Lagrangian \cite{Bronnikov} is given by
\begin{equation}
{\cal L}(\mathcal{F}) =-\frac{{\cal F}}{4\pi\cosh^2\left(a\sqrt[4]{2|{\cal F}|}\right)},
\label{2.2}
\end{equation}
where $a$ is a coupling constant and ${\cal F}=F^{\mu\nu}F_{\mu\nu}/4=(B^2-E^2)/2$ is the field invariant. From action (2.1) one obtains the gravitation and electromagnetic fields equations
\begin{equation}
R_{\mu\nu}-\frac{1}{2}g_{\mu \nu}R+\Lambda g_{\mu \nu} =8\pi G_N T_{\mu \nu},
\label{2.3}
 \end{equation}
\begin{equation}
\partial _{\mu }\left( \sqrt{-g}\mathcal{L}_{\mathcal{F}}F^{\mu \nu}\right)=0.
\label{2.4}
\end{equation}
Here, $R$ and $R_{\mu \nu }$  are the Ricci scalar and tensor, respectively. The stress tensor of electromagnetic fields is given by
\begin{equation}
 T_{\mu\nu }=F_{\mu\rho }F_{\nu }^{~\rho }\mathcal{L}_{\mathcal{F}}+g_{\mu \nu }\mathcal{L}\left( \mathcal{F}\right),
\label{2.5}
\end{equation}
where $\mathcal{L}_{\mathcal{F}}=\partial \mathcal{L}( \mathcal{F})/\partial \mathcal{F}$.
We analyse space-time with the spherical symmetry which has the line element squared
\begin{equation}
ds^{2}=-f(r)dt^{2}+\frac{1}{f(r)}dr^{2}+r^{2}\left( d\theta^{2}+\sin ^{2}\theta d\phi ^{2}\right).
\label{2.6}
\end{equation}
We will consider only magnetized black holes as electrically charged black holes, for NED which has the Maxwell weak-field limit, leads to singularities \cite{Bronnikov}. Then the tensor $F_{\mu\nu}$ possesses the radial electric field $F_{01}=-F_{10}$ and radial
magnetic field $F_{23}=-F_{32}=q\sin(\theta)$, where $q$ is the magnetic charge. The stress tensor is diagonal, $T_{0}^{~0}=T_{r}^{~r}$ and $T_{\theta}^{~\theta}=T_{\phi}^{~\phi}$. One can find the metric function from the relation \cite{Bronnikov}
\begin{equation}
f(r)=1-\frac{2m(r)G_N}{r}.
\label{2.7}
\end{equation}
The mass function is given by
\begin{equation}
m(r)=m_0+4\pi\int_{0}^{r}\rho(r)r^{2}dr,
\label{2.8}
\end{equation}
where the integration constant $m_0$ corresponds to the Schwarzschild mass and $\rho(r)$ is the energy density. It is worth mentioning that $\rho(r)$ in (2.8) includes the term which is due to the cosmological constant.
Here, we study the static magnetic black holes with the field invariant $\mathcal{F}=q^2/(2r^4)$. Thus, the black hole is considered as the magnetic monopole with the magnetic induction field $B=q/r^2$. Making use of Eq. (2.5) we find the magnetic energy density including the term corresponding to the negative cosmological constant
\begin{equation}
\rho=\frac{q}{8\pi r^4\cosh^2(b/r)}-\frac{3}{8\pi G_Nl^2},
\label{2.9}
\end{equation}
where for simplicity we use parameter $b=a\sqrt{q}$.
By virtue of Eqs. (2.8) and (2.9) we obtain the mass function
\begin{equation}
m(r)=m_0+\frac{q^2}{2b}\left[1-\tanh\left(\frac{b}{r}\right)\right]-\frac{r^3}{2G_Nl^2}.
\label{2.10}
\end{equation}
One can find the finite magnetic mass of black holes
\begin{equation}
m_M=\int_0^\infty \frac{qdr}{2r^2\cosh^2(b/r)}=\frac{q^2}{2b}.
\label{2.11}
\end{equation}
Then from Eqs. (2.11) and (2.10) we obtain the mass function
\begin{equation}
m(r)=M-\frac{q^2}{2b}\tanh\left(\frac{b}{r}\right)-\frac{r^3}{2G_Nl^2}.
\label{2.12}
\end{equation}
The ADM black hole mass is given by $M=m_0+m_M$. One can observe that at the Maxwell limit, $b=0$, the magnetic energy is infinite.
With the aid of Eqs. (2.7) and (2.10) one finds the metric function
\begin{equation}
f(r)=1-\frac{2MG_N}{r}+\frac{q^2G_N}{br}\tanh\left(\frac{b}{r}\right)+\frac{r^2}{l^2}.
\label{2.13}
\end{equation}
If we neglect the cosmological constant  $r^2/l^2$ (at $l\rightarrow \infty$), the metric function  Eq. (2.13) as $r\rightarrow \infty$ becomes
\begin{equation}
f(r)=1-\frac{2MG_N}{r}+\frac{q^2G_N}{r^2}-\frac{q^2b^2G_N}{3r^4}+\mathcal{O}(r^{-6})~~~\mbox{as}~r\rightarrow \infty.
\label{2.14}
\end{equation}
Equation (2.14) shows that the correction to the Reissner--Nordstr\"{o}m solution is in the order of $\mathcal{O}(r^{-4})$.
When the Schwarzschild mass is zero, $m_0=0$ (the total black hole mass is the magnetic mass),
we obtain the regular solution from Eq. (2.13) with the asymptotic
\begin{equation}
f(r)=1+\frac{r^2}{l^2}~~~~\mbox{as}~r\rightarrow 0,
\label{2.15}
\end{equation}
so that $f(0)=1$. At $m_0=0$, $G_N=1$, $l=10$ the plots of the metric function (2.13) are depicted in Figs. 1 and 2.
\begin{figure}[h]
\includegraphics [height=3.0in,width=3.0in] {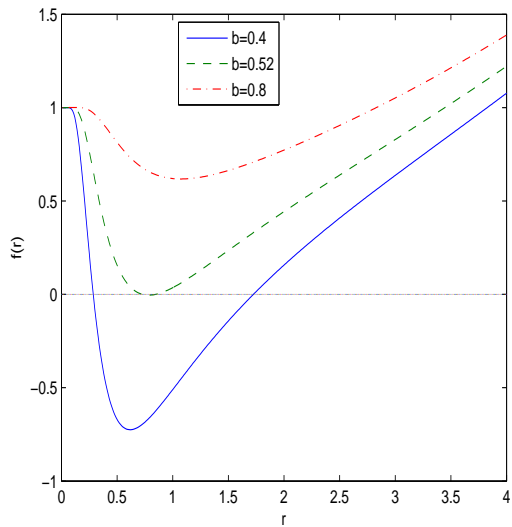}
\caption{\label{fig.1} The plots of the function $f(r)$ at $m_0=0$, $G_N=1$, $l=5$, $q=1$.}
\end{figure}
\begin{figure}[h]
\includegraphics [height=3.0in,width=3.0in] {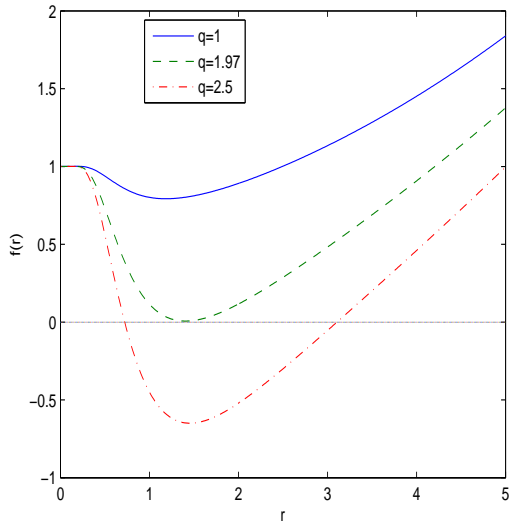}
\caption{\label{fig.2} The plots of the function $f(r)$ at $m_0=0$, $G_N=1$, $l=5$, $b=1$.}
\end{figure}
In accordance with Figs. 1 and 2 black holes may have for various parameters two horizons or one extreme horizon.

\section{First law of black hole thermodynamics and the Smarr relation}

There is an analogy between black hole mechanics and ordinary thermodynamics when black holes are considered as classical objects. Then the surface gravity plays a role of the temperature and event horizon radius is connected with entropy. Hawking argued that black holes emit radiation with a blackbody spectrum \cite{Hawking}. The first law of black hole thermodynamics is given by
$d M=TdS+\Omega dJ+\Phi dq$, where $S$ is an entropy, $M$ is a black hole mass, and $J$ is angular momentum. Here, $M$, $S$, $J$, $q$ are the extensive and $T$, $\Omega$, $\Phi$ are intensive thermodynamic variables. But in this form of first law of black hole thermodynamics the pressure-volume term $Pd V$ is absent. To include the $Pd V$ tern in first law of black hole thermodynamics, a negative cosmological constant $\Lambda$ was considered as the positive vacuum pressure. As a result, the  generalized first law of black hole thermodynamics is formulated as $d M=Td S+ Vd P+\Omega d J+\Phi d q$ where $V=\partial M/\partial P$ (at constant $S$, $J$, $q$) \cite{Kastor, Dolan1, Cvetic}. In this form of the first law of black hole mechanics, $M$ should be interpreted as a chemical enthalpy \cite{Kastor}, $M=U+PV$ with $U$ being an internal energy.

The Smarr relation may be obtained from the first law of black hole thermodynamics by the Euler scaling arguments \cite{Smarr, Kastor}. From the dimensional analysis we obtain units: $[M]=L$, $[S]=L^2$, $[P]=L^{-2}$, $[J]=L^2$, $[q]=L$, $[b]=L$ with $G_N=1$. Here, we use extended phase space and treat $b$ as a thermodynamic variable. Then making use of Euler’s theorem, one obtains the mass function $M(S,P,J,q,b)$,
\begin{equation}
M=2S\frac{\partial M}{\partial S}-2P\frac{\partial M}{\partial P}+2J\frac{\partial M}{\partial J}+q\frac{\partial M}{\partial q}+b\frac{\partial M}{\partial b}.
\label{3.1}
\end{equation}
Here, $\partial M/\partial b\equiv {\cal B}$ is the thermodynamic conjugate to the variable $b$. The black hole volume $V$ and pressure $P$ are given by \cite{Myers, Myers1}
\begin{equation}
V=\frac{4}{3}\pi r_+^3,~~~P=-\frac{\Lambda}{8\pi}=\frac{3}{8\pi l^2}.
\label{3.2}
\end{equation}
In the following we study the non-rotating stationary black hole, and therefore $J=0$. Making use of Eq. (2.13) and the equation for the event horizon radius $r_+$, $f(r_+)=0$, we obtain
\begin{equation}
M=\frac{r_+}{2}+\frac{r_+^3}{2l^2}+\frac{q^2}{2b}\tanh\left(\frac{b}{r_+}\right).
\label{3.3}
\end{equation}
From Eq. (3.3) one finds
\[
d M=\left[\frac{1}{2}+\frac{3r_+^2}{2l^2}
-\frac{q^2}{2r_+^2}\cosh^{-2}\left(\frac{b}{r_+}\right)\right]dr_+-\frac{r_+^3}{l^3}d l
\]
\[
-\left[\frac{q^2}{2b^2}\tanh\left(\frac{b}{r_+}\right)-\frac{q^2}{2br_+}\cosh^{-2}\left(\frac{b}{r_+}\right)\right]d b
\]
\begin{equation}
+\left[\frac{q}{b}\tanh\left(\frac{b}{r_+}\right)\right] dq_.
\label{3.4}
\end{equation}
Making use of expression for Hawking's temperature
\begin{equation}
T=\frac{f'(r)|_{r=r_+}}{4\pi},
\label{3.5}
\end{equation}
and Eqs. (2.13), (3.5) we obtain
\begin{equation}
T=\frac{1}{4\pi}\biggl(\frac{1}{r_+}+\frac{3r_+}{l^2}-\frac{q^2}{r_+^3\cosh^2(b/r_+)}\biggr).
\label{3.6}
\end{equation}
At a particular case $b=0$ ($a=0$), from Eq. (3.6) one finds the Hawking temperature of charged Maxwell-AdS black holes.
By virtue of Eqs. (3.3) and (3.6), we obtain
\begin{equation}
\frac{\partial M(r_+)}{\partial r_+}=2\pi r_+T,
\label{3.7}
\end{equation}
and find the entropy for the NED-AdS black hole
\begin{equation}
S=\int \frac{d M(r_+)}{T}=\int \frac{1}{T}\frac{\partial M(r_+)}{\partial r_+}dr_+=\pi r_+^2.
\label{3.8}
\end{equation}
As a result, we have the same Bekenstein--Hawking entropy as for black holes in the framework of Einstein's gravity. With the help of Eqs. (3.2), (3.3), (3.6)  and (3.8) we obtain the first law of black hole thermodynamics
\begin{equation}
d M = Td S + Vd P + \Phi d q + {\cal B}db.
\label{3.9}
\end{equation}
The magnetic potential $\Phi$ and the thermodynamic conjugate to the variable $b$ are given by
\[
\Phi =\frac{q}{b}\tanh\left(\frac{b}{r_+}\right),~~~~
{\cal B}=-\frac{q^2}{2b^2}\tanh\left(\frac{b}{r_+}\right)
\]
\begin{equation}
+\frac{q^2}{2br_+}\cosh^{-2}\left(\frac{b}{r_+}\right).
\label{3.10}
\end{equation}
The variable ${\cal B}$, which can be considered as vacuum polarization \cite{Mann1}, is important to formulate the Smarr relation. The plots of potential $\Phi$ versus $r_+$ are given in Fig. 3.
\begin{figure}[h]
\includegraphics [height=3.0in,width=3.0in] {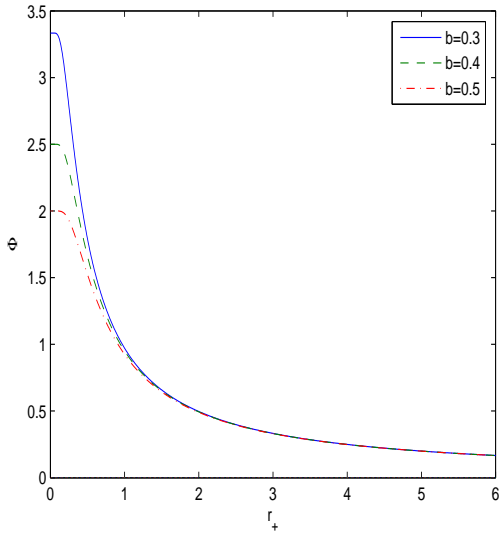}
\caption{\label{fig.3} The plots of the functions $\Phi$ vs. $r_+$ at $q=1$. The solid curve is for $b=0.3$, the dashed curve corresponds to $b=0.4$, and the dashed-doted curve corresponds to $b=0.5$.}
\end{figure}
Figure 3 shows that when variable $b$ increases the magnetic potential decreases and at $r_+\rightarrow \infty$ it becomes zero, $\Phi(\infty)=0$. It is worth mentioning that at $r_+ = 0$ the potential $\Phi$ is the finite value, that is an attractive feature of our model.
As $b\rightarrow 0$ we arrive at the linear electrodynamics and the magnetic potential becomes $\lim_{b\rightarrow 0}\Phi=q/r_+$, i.e. the  potential for the point-like monopole. The plots of the function ${\cal B}$ versus $r_+$ are depicted in Fig. 4.
\begin{figure}[h]
\includegraphics [height=3.0in,width=3.0in] {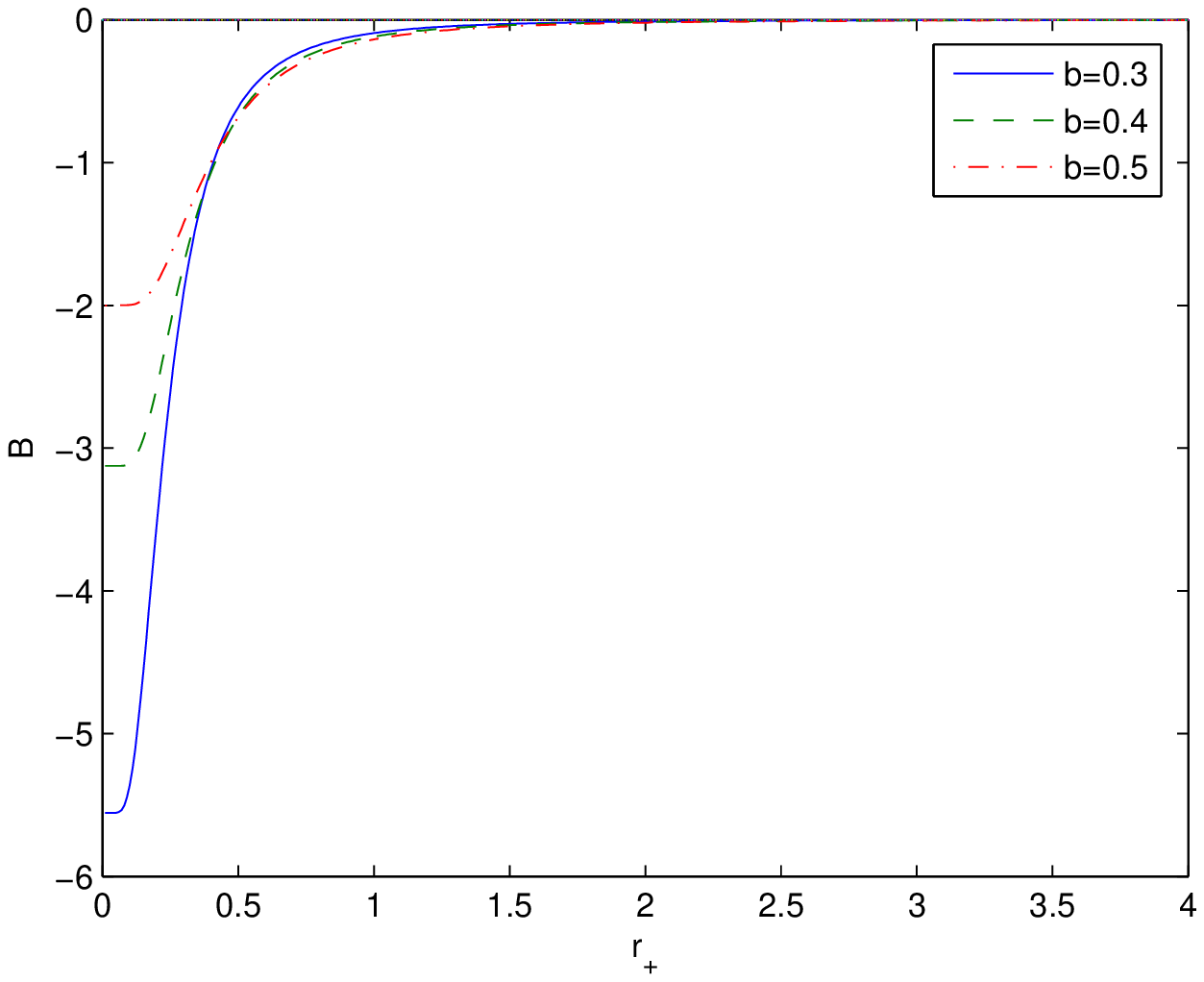}
\caption{\label{fig.4} The plots of the functions  ${\cal B}$ vs. $r_+$ at $q=1$. The solid curve is for $b=0.3$, the dashed curve corresponds to $b=0.4$, and the dashed-doted curve corresponds to $b=0.5$.}
\end{figure}
According to Fig. 4 when variable $b$ increases the $|{\cal B}|$ decreases and as $r_+\rightarrow \infty$ it vanishes, ${\cal B}(\infty)=0$. At $r_+=0$ the vacuum polarization ${\cal B}(0)$ is the finite.
One can check, making use of Eqs. (3.2), (3.3), (3.8) and (3.10), the validity of the generalized Smarr relation
\begin{equation}
M=2ST-2PV+q\Phi+b{\cal B}.
\label{3.11}
\end{equation}
It should be noted that the second law of black hole thermodynamics also holds for NED-AdS space-time (see \cite{Bekenstein, Hawking1}
for Schwarzschild black holes).

\section{Black hole thermodynamics}

With the aid of Eq. (3.6) one finds the equation of state (EoS) for our model of NED-AdS black hole
\begin{equation}
P=\frac{T}{2r_+}-\frac{1}{8\pi r_+^2}+\frac{q^2}{8\pi r_+^4}\cosh^{-2}\left(\frac{b}{r_+}\right).
\label{4.1}
\end{equation}
Setting $b=0$ in Eq. (4.1) we obtain EoS for a charged Maxwell-AdS black hole \cite{Mann2}.
EoS of charged Maxwell-AdS black hole becomes the Van der Waals EoS if the specific volume $v$ is identified
with $2l_Pr_+$. In our units $l_P=\sqrt{G_N}=1$ and the horizon diameter $2r_+$ mimics the specific volume of the Van der Waals fluid.  Then, Eq. (4.1) is rewritten as
\begin{equation}
P=\frac{T}{v}-\frac{1}{2\pi v^2}+\frac{2q^2}{\pi v^4}\cosh^{-2}\left(\frac{2b}{v}\right).
\label{4.2}
\end{equation}
EoS (4.2) is qualitatively similar to EoS of the Van der Waals fluid. Critical points corresponding to possible phase transitions can be obtained by using equations for the inflection points in the $P-v$ diagram,
\[
\frac{\partial P}{\partial v}=-\frac{T}{v^2}+\frac{1}{\pi v^3}-\frac{8q^2}{\pi v^5}\cosh^{-2}\left(\frac{2b}{v}\right)
\]
\[
+\frac{8q^2b}{\pi v^6}\frac{\tanh(2b/v)}{\cosh^2(2b/v)}=0,
\]
\[
\frac{\partial^2 P}{\partial v^2}=\frac{2T}{v^3}-\frac{3}{\pi v^4}+\frac{40q^2}{\pi v^6}\cosh^{-2}\left(\frac{2b}{v}\right)
-\frac{80q^2b}{\pi v^7}\frac{\tanh(2b/v)}{\cosh^2(2b/v)}
\]
\[
-\frac{16q^2b^2}{\pi v^8}\cosh^{-2}\left(\frac{2b}{v}\right)\biggl[\cosh^{-2}\left(\frac{2b}{v}\right)
\]
\begin{equation}
-2\tanh^2\left(\frac{2b}{v}\right)\biggr]=0.
\label{4.3}
\end{equation}
Making use of the system of equations (4.3), we obtain the critical points equation
\[
v_c^4-8q^2\cosh^{-2}\left(\frac{2b}{v_c}\right)\biggl\{3v_c^2-2b\biggl[4v_c\tanh\left(\frac{2b}{v_c}\right)
\]
\begin{equation}
+b\cosh^{-2}\left(\frac{2b}{v_c}\right)-2b\tanh^2\left(\frac{2b}{v_c}\right)\biggr]\biggr\}=0.
\label{4.4}
\end{equation}
With the help of Eq. (4.3) one finds the critical temperature
\begin{equation}
T_c=\frac{1}{\pi v_c}-\frac{8q^2}{\pi v_c^3}\cosh^{-2}\left(\frac{2b}{v_c}\right)
+\frac{8q^2b}{\pi v_c^4}\frac{\tanh(2b/v_c)}{\cosh^2(2b/v_c)}.
\label{4.5}
\end{equation}
By virtue of Eqs. (4.2) and (4.5) we obtain the critical pressure
\begin{equation}
P_c=\frac{1}{2\pi v_c^2}-\frac{6q^2}{\pi v_c^4}\cosh^{-2}\left(\frac{2b}{v_c}\right)
+\frac{8q^2b}{\pi v_c^5}\frac{\tanh(2b/v_c)}{\cosh^2(2b/v_c)}.
\label{4.6}
\end{equation}
Approximate solutions to Eq. (4.4) and critical temperatures and pressures for various $q$ values at $b=0.001$ are given in Table I.
\begin{table}[ht]
\caption{Critical values of the specific volume, temperatures and pressures at $b=0.001$}
\centering
\begin{tabular}{c c c c c c c c c c}\\[1ex]
\hline
$q $ & 1 & 1.1 & 1.2 & 1.3 & 1.4 & 1.5 & 1.6 & 1.7     \\[0.5ex]
\hline
$v_c$ &4.90 & 5.39 & 5.88 & 6.37 & 6.86 & 7.51 & 7.84 & 8.33   \\[0.5ex]
\hline
$T_c$ &0.0433 & 0.0394 & 0.0361 & 0.0333 & 0.0309 & 0.0289 & 0.0271 & 0.0255 \\[0.5ex]
\hline
$P_c$ &0.0033&0.0027&0.0023&0.0020&0.0017&0.0015&0.0013 &0.0011 \\[0.5ex]
\hline
\end{tabular}
\end{table}
At the point $v_c$  second-order phase transition occurs. The $P-v$ diagrams are depicted in Figs. 5 and 6 for various values of Hawking temperature $T$.
\begin{figure}[h]
\includegraphics [height=3.0in,width=3.0in] {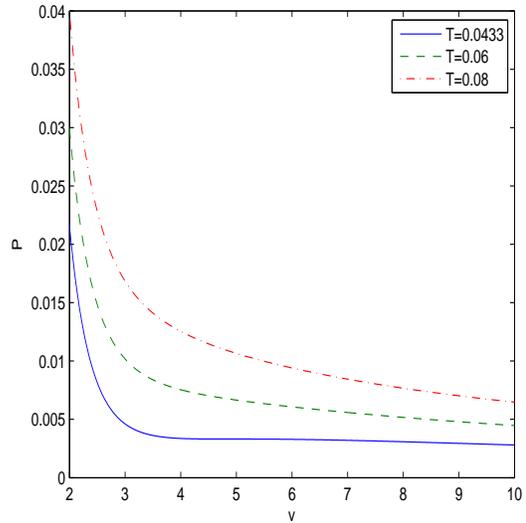}
\caption{\label{fig.5} The plots of the function $P$ vs. $v$ at $q=1$, $b=0.001$. The critical isotherm corresponds to $T_{c}=0.0433$.}
\end{figure}
\begin{figure}[h]
\includegraphics [height=3.0in,width=3.0in] {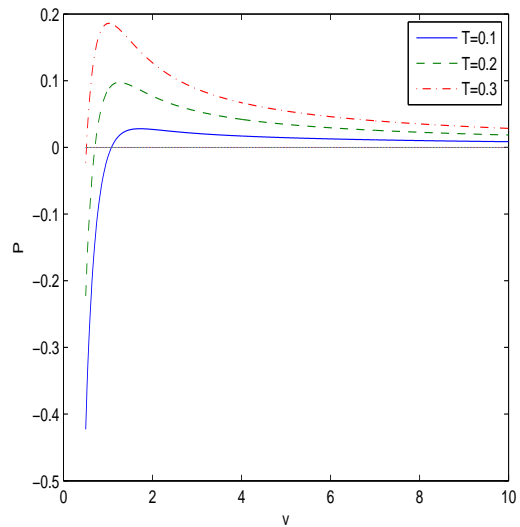}
\caption{\label{fig.6} The plots of the function $P$ vs. $v$ at $q=b=1$. The non-critical isotherms correspond to $T=0.1$, $0.2$ and
$0.3$.}
\end{figure}
Table I (see also Fig. 5) shows that for $q=1$, $b=0.001$ there is the critical specific volume $v_{c}\approx 4.90$ ($T_c=0.0433$). This inflection point is shown in Fig. 5 for $T=0.0433$ ($v_c\approx 4.9$) where second-order phase transition occurs. Non-critical behaviour of $P-v$ diagrams for $T=0.1$, $0.2$, $0.3$ and $q=b=1$ is depicted in Fig. 6. For $q=b=1$ there are not real solutions to Eq. (4.4) and critical points, and therefore, only first-order phase transitions hold. It is worth noting that at $b=0$ ($a=0$) one comes to Maxwell electrodynamics coupled to AdS-Einstein's gravity and there is a critical point showing the possibility of second-order phase transition. This case was studied in \cite{Teo,Mann1}. According to Eqs. (4.4)-(4.6) critical points depend on $b$. With the help of Eqs. (4.5) and (4.6) we obtain the critical ratio
\[
\rho_c=\frac{P_cv_c}{T_c}
\]
\begin{equation}
=\frac{v_c^3\cosh^2(2b/v_c)-12v_cq^2+16q^2b\tanh(2b/v_c)}{2[v_c^3\cosh^2(2b/v_c)-8v_cq^2+8q^2b\tanh(2b/v_c)]},
\label{4.7}
\end{equation}
It follows from Eq. (4.4) that at $b=0$, $v_c=24q^2$ and then from (4.7) one gets $\rho_c=3/8$ which is the same as for a Van der Waals fluid.
At $b\neq 0$ we have $\rho_c\neq 3/8$ and the black hole critical behaviour is different from a Van der Waals fluid.

\subsection{ The Gibbs free energy}

To study global black hole stability we calculate the Gibbs free energy for a fixed charge, variable $b$ and pressure
\begin{equation}
G=M-TS.
\label{4.8}
\end{equation}
We treat mass $M$ as a chemical enthalpy, which is the total energy of a system with its internal energy $U$ and the energy $PV$ to displace the vacuum energy of its environment, $M =U+PV$. Making use of Eqs. (3.3) and (4.8) one finds
\begin{equation}
G=\frac{r_+}{4}-\frac{2\pi r_+^3P}{3}+\frac{q^2}{2b}\tanh\left(\frac{b}{r_+}\right)+\frac{q^2}{4r_+\cosh^2(b/r_+)}.
\label{4.9}
\end{equation}
The plots of $G$ versus $T$ are given in Fig. 7, where we took onto account that $r_+$, according to Eq. (4.1), is a function of $P$ and $T$.
\begin{figure}[h]
\includegraphics [height=3.0in,width=3.0in] {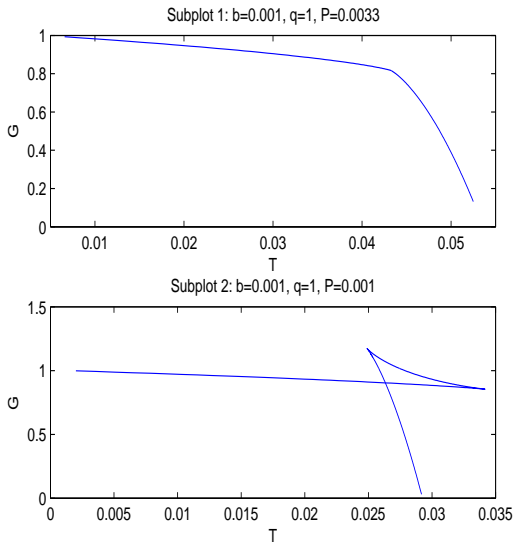}
\caption{\label{fig.7} The plots of the Gibbs free energy $G$ vs. $T$ with $b=0.001$, $q=1$.}
\end{figure}
The Gibbs free energy $G$ behaviour depends on pressure $P$ and coupling $a$ ($b=a\sqrt{q}$). Here, in Fig. 7, we consider the example with $q=1$, $b=0.001$ where there is the critical point $v_c\approx 4.9$, $T_c\approx 0.0433$, $P_c\approx 0.0033$.
The behaviour of the Gibbs free energy is similar to the charged RN-AdS black hole with one critical point and the corresponding first-order phase transition between small and large black holes.
The Gibbs free energy shows 'swallowtail' behaviour with the first-order phase transition between two branches for $P<P_c$ (Subplots 2 in Fig. 7). At the point $P_c$, $T_c$ second-order phase transition occurs and there is no first-order phase transition in the system for $P>P_c$.

\section{Joule--Thomson expansion of NED-AdS black holes}

The Joule--Thomson black hole expansion is isenthalpic adiabatic expansion with black hole mass $M$ treated as the enthalpy which is constant during the expansion. To study cooling-heating phases one introduces the Joule--Thomson thermodynamic coefficient which is given by
\begin{equation}
\mu_J=\left(\frac{\partial T}{\partial P}\right)_M=\frac{1}{C_P}\left[ T\left(\frac{\partial V}{\partial T}\right)_P-V\right]=\frac{(\partial T/\partial r_+)_M}{(\partial P/\partial r_+)_M},
\label{5.1}
\end{equation}
that is the slope in the $P-T$ diagram. When $\mu_J>0$ we have cooling process during the expansion while at $\mu_J<0$ the heating process occurs. At the inversion temperature $T_i$ ($\mu_J(T_i)=0$) the sign of $\mu_J$ is changed. When during the expansion the initial temperature is higher than inversion temperature $T_i$, we have the cooling phase, $\mu_J>0$, and the final temperature decreases. If the initial temperature is lower than $T_i$, $\mu_J<0$, then the final temperature increases and the heating  phase takes place. Making use of Eq. (5.1) and equation for the inversion temperature $\mu_J(T_i)=0$, we find
\begin{equation}
T_i=V\left(\frac{\partial T}{\partial V}\right)_P=\frac{r_+}{3}\left(\frac{\partial T}{\partial r_+}\right)_P.
\label{5.2}
\end{equation}
The inversion temperature represents a borderline between cooling and heating processes. The line of inversion temperature
crosses a point in maximum of the $P-T$ diagram where its slope is changed and this point separates cooling and heating black hole phases
\cite{Yaraie, Mo, Rizwan}. Black hole equation of  state (3.6) may be represented in the form
\begin{equation}
T=\frac{1}{4\pi r_+}+2P r_+-\frac{q^2}{4\pi r_+^3}\cosh^{-2}\left(\frac{b}{r_+}\right).
\label{5.3}
\end{equation}
From Eq. (3.3) and the equation for pressure $P=3/(8\pi l^2)$ we obtain
\begin{equation}
P=\frac{3M}{4\pi r_+^3}-\frac{3}{8\pi r_+^2}-\frac{3 q^2}{8\pi br_+^3}\tanh\left(\frac{b}{r_+}\right).
\label{5.4}
\end{equation}
Equations (5.3) and (5.4) are the parametric form of the $P-T$ diagram which is depicted in Fig. 8.
\begin{figure}[h]
\includegraphics [height=3.0in,width=3.0in]{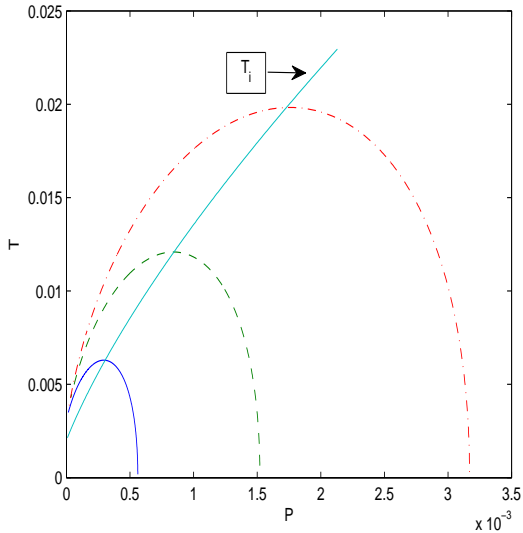}
\caption{\label{fig.8} The plots of the temperature $T$ versus pressure $P$  and the inversion temperature $T_i$ with $q=10$, $a=1$ ($b=a\sqrt{q}$). The $P-T$ diagram corresponds to masses $M=11,12,13$ from bottom to top.}
\end{figure}
 Making use of Eqs. (5.2) and (5.3) one obtains the inversion pressure
\[
P_i=\frac{3q^2}{16\pi r_+^4}\left[1+\cosh^{-2}\left(\frac{b}{r_+}\right)\left(1-\frac{2b}{3r_+}\tanh\left(\frac{b}{r_+}\right)\right)\right]
\]
\begin{equation}
-\frac{1}{4\pi r_+^2}.
\label{5.5}
\end{equation}
From Eqs. (5.3) and (5.5) we find the inversion temperature
\[
T_i=\frac{q^2}{8\pi r_+^4}\left[3r_++\cosh^{-2}\left(\frac{b}{r_+}\right)\left(r_+-2b\tanh\left(\frac{b}{r_+}\right)\right)\right]
\]
\begin{equation}
-\frac{1}{4\pi r_+}.
\label{5.6}
\end{equation}
Approximate solutions to equation $P_i=0$ (following from Eq. (5.5)) and the minimum of inversion temperature at $q=1$ are given in Table II.
\begin{table}[ht]
\caption{Minimum of event horizon radius corresponding to minimum of inversion temperature at $q=1$}
\centering
\begin{tabular}{c c c c c c c c c c c}\\[1ex]
\hline
$b$ & 0.9 & 0.8 & 0.7 & 0.6 & 0.5 & 0.4 & 0.3    \\[0.5ex]
\hline
$r_i^{min}$ &0.971 & 1.014 & 1.059 & 1.101 & 1.139 & 1.170 & 1.194   \\[0.5ex]
\hline
$T_i^{min}$ &0.0413 & 0.0352 & 0.0306 & 0.0275 & 0.0252 & 0.0237 & 0.0227  \\[0.5ex]
\hline
\end{tabular}
\end{table}
According to Table II, when coupling $a$ ($b=a\sqrt{q}$) decreases the minimum inversion temperature decreases.
Making use of Eqs. (5.5) and (5.6), at $b=0$ ($a=0$), one obtains the minimum inversion temperature $T_i^{min}$ and corresponding event horizon radius
\begin{equation}
T_i^{min}=\frac{1}{6\sqrt{6}\pi q}, ~~~~r_+^{min}=\frac{\sqrt{6}q}{2}.
\label{5.7}
\end{equation}
With the help of Eqs. (4.4) and (4.5) at $b=0$ we obtain $v_c=2\sqrt{6}q$, $T_c=1/(3\sqrt{6}\pi q)$ and making use of Eq. (5.7) one finds $T_i^{min}=T_c/2$ that is in accordance with the result obtained in \cite{Aydiner} for electrically charged Maxwell-AdS black holes. Equations (5.5) and (5.6) represent inversion temperature $T_i$ versus $P_i$ in the parametric form that is depicted in Fig. 9. In accordance with Fig. 8 the inversion point increases with increasing the mass of black holes. The plots of the inversion curve $P_i-T_i$ with various parameters are depicted in Fig. 9.
\begin{figure}[h]
\includegraphics [height=3.0in,width=3.0in] {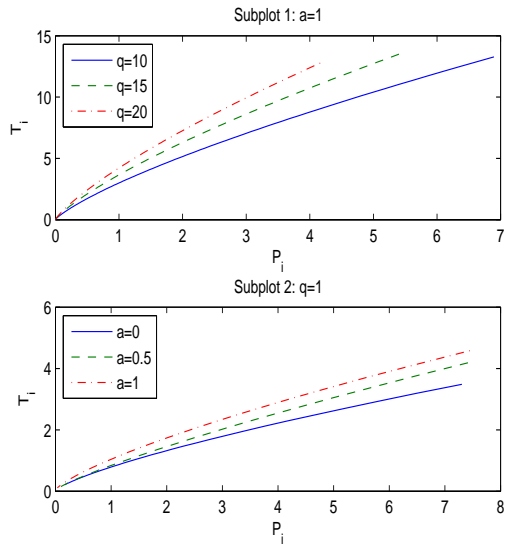}
\caption{\label{fig.9} The plots of the inversion temperature $T_i$ versus pressure $P_i$ for $q=10$, $15$ and $20$, $a=1$ and for $a=0$, $0.5$ and $1$, $q=1$.}
\end{figure}
According to Fig. 9, Subplot 1, with increasing magnetic charge $q$, for fixed black hole coupling $a$, the inversion temperature increases.
Figure  9, Subplot 2, shows that if the coupling $a$ increases, at fixed magnetic charge $q$, the inversion temperature also increases.
With the aid of Eqs. (5.1), (5.3) and (5.4) we find the Joule--Thomson coefficient
\[
\mu_J=\frac{4\left[br_+^2(r_+-6M)+3q^2r_+^2t+bq^2c(3r_+-bt\right]}{3\left[2br_+(r_+-3M)+3q^2r_+t
+q^2bc\right]},
\]
\begin{equation}
c\equiv\cosh^{-2}\left(\frac{b}{r_+}\right),~~~~t\equiv\tanh\left(\frac{b}{r_+}\right).
\label{5.8}
\end{equation}
In the case when the Joule--Thomson coefficient is positive, $\mu_J>0$, we have a cooling process, while when $\mu_J<0$, a heating process occurs. The area in Fig. 8 with $\mu_J>0$ is in the left side of the borderline of inversion temperature, and $\mu_J<0$ takes place in the right side of the borderline $T_i$.

\section{Conclusion and summary}

We have studied NED-AdS magnetic black hole, and found metric and mass functions and their asymptotic. The black holes can have two horizons or one extreme horizon depending on the model parameters: coupling $a$, magnetic charge $q$, and the AdS radius $l$. Corrections, in the order of ${\cal O}(r^{-4})$, to the Reissner--Nordstr\"{o}m solution have been obtained. It was shown that when NED parameter $b$ increases the event horizon radius $r_+$ decreases, whereas if magnetic charge increases the event horizon radius increases. We have studied
thermodynamics and pase transitions of NED-AdS black holes in an extended thermodynamic phase space where the cosmological constant is treated as a thermodynamic pressure and the mass of the black hole being the chemical enthalpy. We have introduced a thermodynamic quantity (vacuum polarization) ${\cal B}$ conjugated to NED parameter $b$ and magnetic potential $\Phi$ conjugated to charge $q$. The attractive feature of potential $\Phi$ is that it is finite at $r_+=0$ while in Maxwell electrodynamics the potential $q/r$ is singular. This is because singularities
are smoothed by the coupling $a$ ($b=a\sqrt{q}$). When $a=0$ NED becomes Maxwell electrodynamics. We have formulated
the first law of black hole thermodynamics in an extended phase space and it was demonstrated that the generalized Smarr relation holds. In this picture the negative cosmological constant plays a role of the pressure. There is an analogy of thermodynamics considered with the Van der Walls liquid–gas system. The Gibbs free energy have been calculated and critical temperature and pressure have been found. It was demonstrated that phase transitions in black holes at some conditions occur. We have obtained the critical ratio $\rho_c=3/8+{\cal O}(b)$ wearies the Van der Waals critical ratio is $3/8$. Cooling and heating phases of NED-AdS black holes via the Joule--Thomson adiabatic expansion have been studied. We have calculated the Joule--Thomson thermodynamic coefficient $\mu_J$ and inversion temperature $T_i$ which allow us to study cooling-heating phases. We have cooling  and heating processes, during the Joule--Thomson adiabatic expansion, at $\mu_J>0$ and $\mu_J<0$, respectively. The inversion temperature gives a borderline between cooling and heating processes and separates the isenthalpic plots into two branches.

Although the black hole thermodynamics mimics the Van der Waals fluid there are many open questions to be answered.

\end{document}